\documentclass{article}

\usepackage[utf8]{inputenc} 
\usepackage[T1]{fontenc}    
\usepackage{hyperref}       
\usepackage{url}            
\usepackage{booktabs}       
\usepackage{amsfonts}       
\usepackage{nicefrac}       
\usepackage{microtype}      
\usepackage{graphicx}
\usepackage{multirow}
\usepackage{array}
\usepackage{diagbox}
\usepackage{subfigure}
\usepackage{xcolor}
\usepackage{authblk}
\usepackage{subfigure}
\title{Multi-label Detection and Classification of\\Red Blood Cells in Microscopic Images}

\author[1]{Wei Qiu*}
\author[1]{Jiaming Guo*}
\author[2]{Xiang Li*}
\author[1]{Mengjia Xu}
\author[1]{Mo Zhang}
\author[2]{Ning Guo}
\author[2]{Quanzheng Li}
\affil[1]{Peking University}
\affil[2]{Department of Radiology, Massachusetts General Hospital}

\date{}

\begin{document}

\maketitle

\begin{abstract}
Cell detection and cell type classification from biomedical images play an important role for high-throughput imaging and various clinical application. While classification of single cell sample can be performed with standard computer vision and machine learning methods, analysis of multi-label samples (region containing congregating cells) is more challenging, as separation of individual cells can be difficult (e.g. touching cells) or even impossible (e.g. overlapping cells). As multi-instance images are common in analyzing Red Blood Cell (RBC) for Sickle Cell Disease (SCD) diagnosis, we develop and implement a multi-instance cell detection and classification framework to address this challenge. The framework firstly trains a region proposal model based on Region-based Convolutional Network (RCNN) to obtain bounding-boxes of regions potentially containing single or multiple cells from input microscopic images, which are extracted as image patches. High-level image features are then calculated from image patches through a pre-trained Convolutional Neural Network (CNN) with ResNet-50 structure. Using these image features inputs, six networks are then trained to make multi-label prediction of whether a given patch contains cells belonging to a specific cell type. As the six networks are trained with image patches consisting of both individual cells and touching/overlapping cells, they can effectively recognize cell types that are presented in multi-instance image samples. Finally, for the purpose of SCD testing, we train another machine learning classifier to predict whether the given image patch contains abnormal cell type based on outputs from the six networks. Testing result of the proposed framework shows that it can achieve good performance in automatic cell detection and classification.
\end{abstract}

\section{Introduction}
\label{sec:intro}
Sickle cell disease (SCD) is a type of inherited red blood cell (RBC) disorder which can cause life-threatening complications. Automatic classification and diseased cell detection based on cell texture and morphological features have become a viable and important approach for SCD diagnosis, as manual inspection of RBC images is time and labor-consuming. More generally, automatic cell detection and cell type classification is a crucial step of high-throughput imaging as well as many other clinical applications. Towards the purpose of cell detection and classification, various solutions have been developed, such as CellProfiler \cite{carpenter2006cellprofiler}, CellTrack \cite{sacan2008celltrack} or Fiji\cite{schindelin2012fiji}. Recent advancement of deep learning-based approaches \cite{Shirazi2017Extreme, Xu2017A, DBLP:journals/corr/abs-1708-03307, hirimutugoda2010image, Elsalamony2016Healthy, zhang2018rbc, Xie18} has shown superior performance in extracting more discriminative image features with higher generalizability for various biomedical image analysis tasks including cell classification, detection, semantic segmentation and counting.

While deep learning-based approaches have achieved good performance in classifying single cell patches \cite{Win_2018, liang2018combining, gao2017hep}, in practice a common challenge is the presence of multiple cells congregating together in one sample image patch. We formulate this challenge as the “multi-label classification” problem, where it can be difficult (e.g. touching cells) or even impossible (e.g. overlapping cells) to fully separate individual instances out in those samples. As normal classifiers are trained for only dealing with a single instance, those multi-label samples have to be discarded \cite{Xu2017A} during training, and can cause incorrect classification results if such samples are presented in testing data. However, touching cells and overlapping cells are very common in microscopic images, so it is significant to solve this multi-label classification problem. Among various multi-instance methods that have been previously developed, CapsNet \cite{sabour2017dynamic} can analyze highly overlapping objects and has inspired many applications based on it. However, most of the current models developed using CapsNet are focusing on single-label classification problem \cite{chen2018learning, de2018capsule, iesmantas2018convolutional}, due to the limitation in the original CapsNet that it does not allow more than one instance of the same class to be presented in the image. Since there are many patches including multiple cells from the same class congregating together, CapsNet can’t deal with multi-label RBCs classification problem.

To address the challenge of multi-label classification in biomedical image analysis, while at the same time aiming at improving the diagnostic accuracy and efficiency for SCD, we propose a cell detection and classification framework that can automatically extract image patches consisting of single or multiple cells, and perform multi-label classification as well as abnormal cell detection on the extracted image patches. The proposed framework including three steps. Firstly, we applied a Faster-RCNN to automatically extract single-cell and multi-cell patches from a complete microscopic image. Secondly, we implemented a pre-trained ResNet for feature extraction and developed multiple networks to obtain the predicted cell types in the patches. Finally, we exploited a Gradient Boosting Classifier to determine the presence of “abnormal” cell types in a given cell patch. 

Our approach makes the following main contributions. Firstly, we implement a deep learning framework Faster-RCNN on RBC microscopic images for cell detection. What’s more, we propose a simple but effective multi-label classification method that is able to classify single cell patches and multiple cells patches together.  The proposed approach is trained and tested on whole microscopic images. The high accuracy for both detection and multi-label classification demonstrates the effectiveness of the proposed framework for the automatic detection and multi-label classification of RBCs. To the best of our knowledge, this is the first work attempting to solve multi-label classification problem of RBCs.

The rest of this paper is organized as follows: in Section~\ref{section:method} we present our materials and methods and in Section~\ref{section:result} we present the experimental results, a comparative analysis, and a discussion. Finally, in Section~\ref{section:conclusion} we present the conclusion and discussion.

\section{Material and methods}
\label{section:method}
\subsection{Data acquisition and approach overview}

\begin{figure}[ht]
\centering
\includegraphics[width=1.0\textwidth]{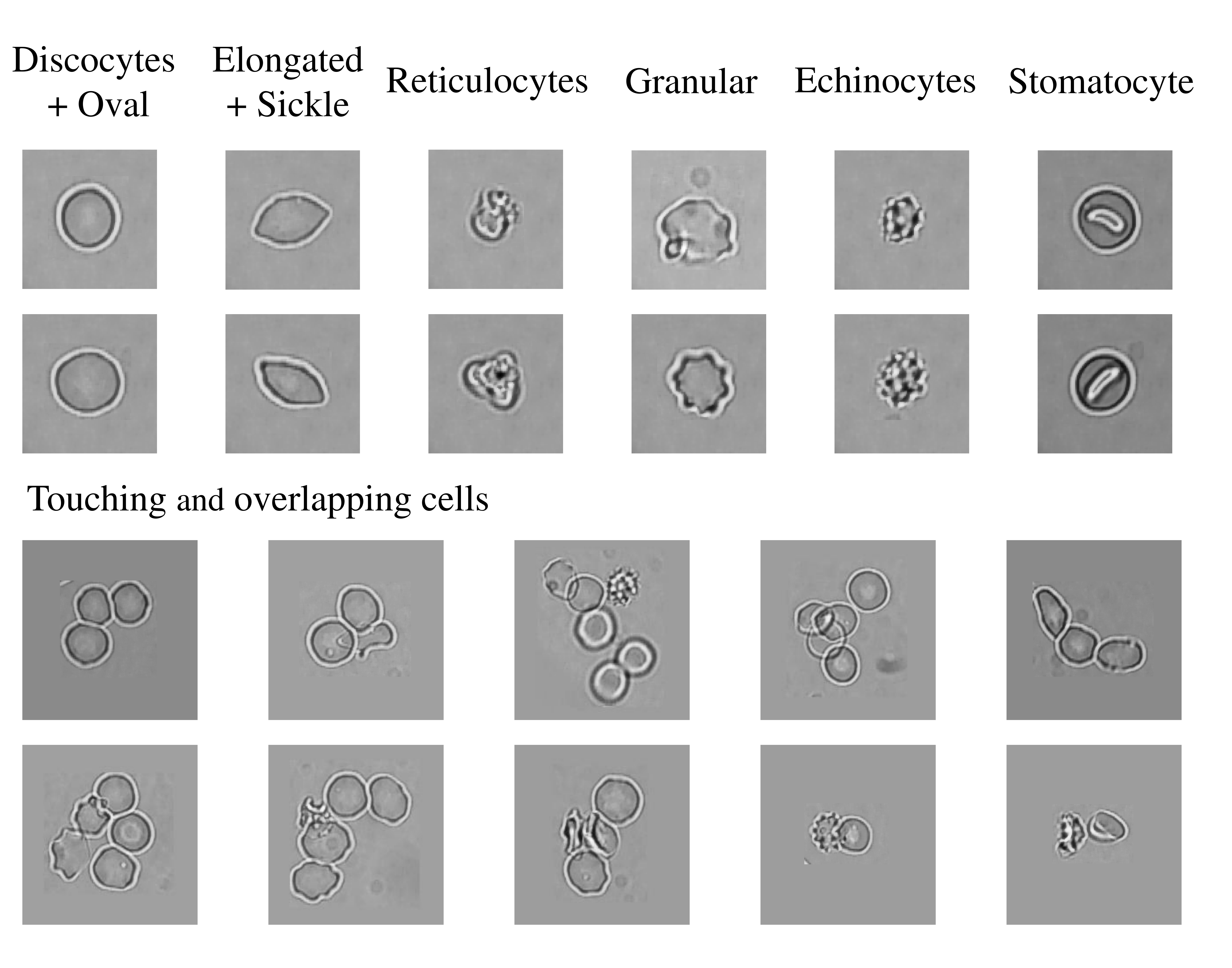}
\caption{Top panel: Sample image patches belonging to six cell types. \protect\\
Bottom panel: Sample images patches containing touching/overlapping cells.}
\label{figure:sample of cells}
\end{figure}

RBC microscopic images used in this work are collected from UPMC (University of Pittsburgh Medical Center) and Massachusetts General Hospital. Raw data contains 313 images with the size of 1920$\times$1080. Details of data acquisition and description can be found in \cite{Xu2017A}. Data used in this work includes 1080 single-cell patches processed in \cite{Xu2017A}, as well as 1389 multi-cell patches with touching/overlapping cells that are manually identified from raw images. According to the protocol in \cite{Xu2017A}, we define six cell types for RBC, visualizations of the six cell types as well as samples of touching/overlapping cells can be found in Figure~\ref{figure:sample of cells}.

\begin{figure}[ht]
\centering
\includegraphics[width=1.0\textwidth]{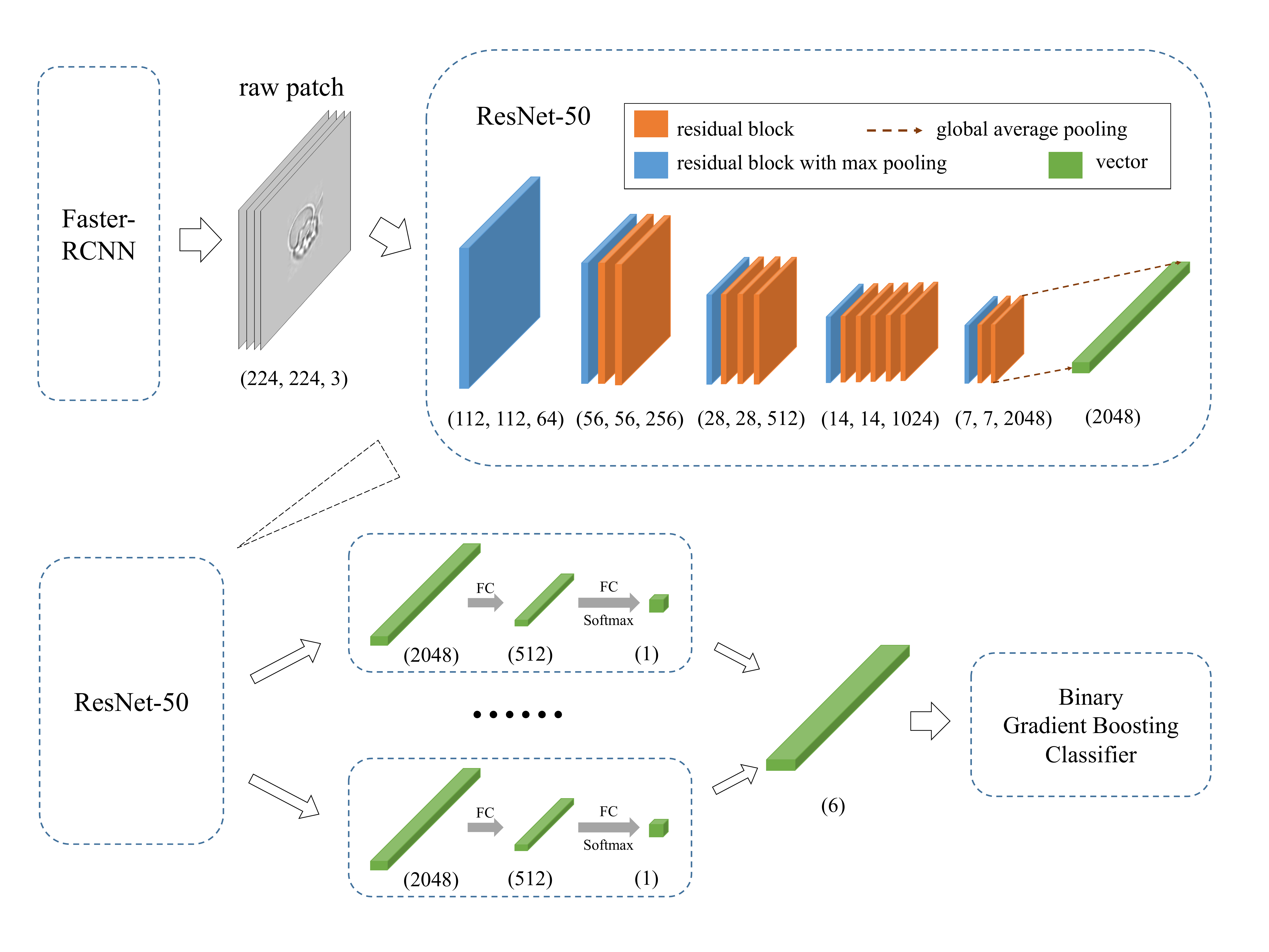}
\caption{Workflow of the cell detection and multi-label classification method.}
\label{figure:workflow}
\end{figure}

Figure~\ref{figure:workflow} illustrates the workflow of our approach, which involves cell detection and multi-label classification. The framework firstly performs region proposal of the full-scale microscopic input image through a Region-based Convolutional Network (RCNN) implemented by Faster-RCNN \cite{ren2015faster} and extract image patches automatically. In the next step, the proposed framework uses Convolutional Neural Network (CNN) with network structure of ResNet-50 and pre-trained on ImageNet dataset \cite{deng2009imagenet} to extract high-level image features (i.e. outputs from the last convolution layer) from the image patches. Afterwards, six classification networks using the extracted image features as input are trained to classify whether the input image patch contains cell(s) belonging to a specific cell type or not. A similar scheme for multi-label classification has also been applied in previous works \cite{zhu2017multi, sarafianos2017curriculum, coudray2018classification}. For the purpose of SCD testing, we further apply Gradient Boosting Classifier to determine whether the given image patch contains “abnormal” cell types or not, based on the outputs from six classification networks. The proposed framework is tested on microscopic RBC images from SCD patients, showing its capability of performing fully automatic cell detection, cell type classification and SCD testing.

\subsection{Cell detection with Faster-RCNN}
In order to automatically extract image patches from the full-scale 1920$\times$1080 microscopic images, we utilize Faster-RCNN \cite{ren2015faster} model which has achieved state-of-the-art performances and high process speed for object detection and region proposal tasks. We modified Faster-RCNN with a ResNet-101 network for better feature extraction performance. 

\begin{figure}[ht]
\centering
\subfigure[image]{
\begin{minipage}{0.4\textwidth}
\centering
\includegraphics[width=1\textwidth]{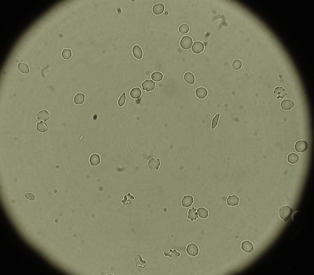}
\end{minipage}
}
\subfigure[label]{
\begin{minipage}{0.4\textwidth}
\centering
\includegraphics[width=1\textwidth]{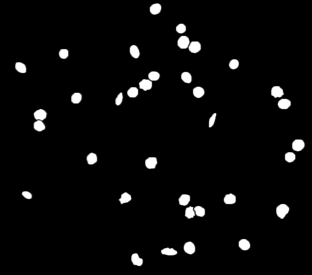}
\end{minipage}
}
\caption{A selected image and its corresponding label.}
\label{figure:image and label}
\end{figure}

To generate the input of Faster-RCNN, we obtain bounding boxes from ground truth cell position labels (Figure~\ref{figure:image and label}) through BFS (Breadth First Search) algorithm. In the labels, background pixels are black, where cell pixels are white. BFS finds out every connected region of white cell pixels in each ground truth label, be it a single-cell region or a multi-cell region, and uses a bounding box to limit each region of pixels.

\begin{figure}[ht]
\centering
\includegraphics[width=0.9\textwidth]{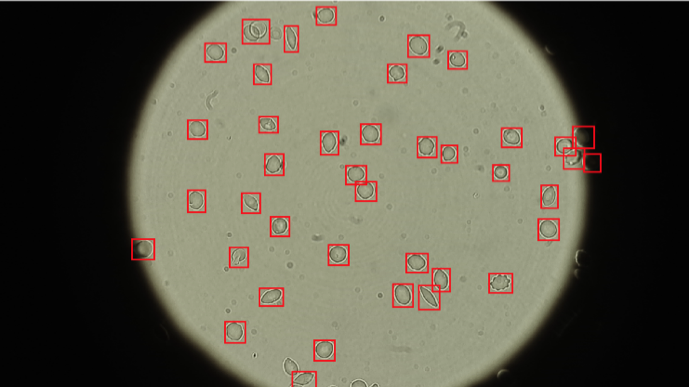}
\caption{Selected detection results from faster-RCNN.}
\label{figure:detection results}
\end{figure}

In the proposed framework, the ResNet-101 network is pre-trained on ImageNet, and is trained on whole microscopic images (with bounding boxes) in order to extract patches containing single or multiple cells. Optimization of the detection process is performed by Momentum Optimizer with learning rate $10^{-3}$, decay $10^{-6}$, momentum 0.9, batch size (10) and epoch (1000). The training process converges in 1 hour on a Linux PC with 8G RAM and a GTX 1070 GPU, and the detection for each test image takes less than 0.5 seconds on the same device. The Faster-RCNN stage of our network can accurately detect multi-instance image patches. Selected detection results are shown in Figure~\ref{figure:detection results}. After the detection, extracted image patches are resized to 224$\times$224 pixels in order to be used as input for later networks.

\subsection{Multi-label classification with transfer learning}
\begin{figure}[ht]
\centering
\includegraphics[width=1.0\textwidth]{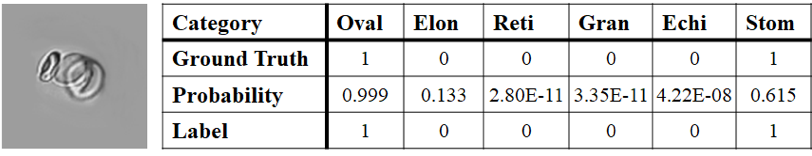}
\caption{Sample output of six binary classification networks.  The table in the right shows ground truth, predicting probability and predicting label of the left cell patch generated by the proposed multi-label classification networks.}
\label{figure:sample of classification output}
\end{figure}

In order to perform effective multi-label cell classification in a supervised approach, one major challenge to overcome is the lack of training samples, which is a common problem when applying deep learning methods for medical image analysis \cite{hoo2016deep, Phan2016Transfer, cheng2017transfer, Vesal2018Classification}. To solve this, we develop a transfer learning scheme which utilizes ResNet-50 network \cite{He_2016_CVPR} pre-trained on ImageNet \cite{deng2009imagenet} to extract high-level image features. Specifically, the pre-trained ResNet-50 is applied to all the available sample image patches (i.e. using them as testing input). The ResNet part in Figure~\ref{figure:workflow} shows the architecture of ResNet-50. Outputs from the last convolution layer of ResNet-50, which can be considered as a high-level representation of the input image patches, are then stored and used for training the later cell type classification network. In this way, we transfer the massive information in the ImageNet database to this application through convolution operations, resulting in the extracted image features. These image features, formed as a 2048-d vector for each input image patch, where 2048 is the number of convolution kernels in the last convolution layer of ResNet-50. The framework then trains six customized fully connected networks with a 512-d fully-connected layer, a 1-d fully-connected layer and a softmax output layer. Each network performs binary classification for one cell type, where its input is the 2048-d feature vector, and output is a decimal range from 0 to 1 which indicates the probability of whether a certain cell type is presented in the input image patch. If the output is greater than 0.5, we consider that the input image patch contains that certain type of cells. Optimization of the classification networks is performed by Adam optimizer \cite{Kingma2014Adam} with a learning rate of $10^{-5}$. The loss is measured by cross-entropy with L2 regularization. Finally, outputs of the six networks are aggregated together into a 6-d vector, showing the probability for each of the six cell types. The predicted cell types can be generated by a threshold of 0.5. It should be noted that this output vector is not normalized (i.e. the sum of probability is not 1), as we allow more than one cell types presented in the given image patch. Figure~\ref{figure:sample of classification output} illustrates a sample output of the proposed multi-label classification method.

\subsection{Binary classification Gradient Boosting Classifier}

As the ultimate goal of RBC image analysis for SCD testing is the detection of whether abnormal cells are presented in the given microscopic image, where “abnormal” is defined by five cell types: “Elongated and Sickle”, “Reticulocytes”, “Granular”, “Echinocytes” and “Stomatocyte”, we further construct a binary classifier using Gradient Boosting Classifier \cite{friedman2001greedy} to discriminate “normal” cells versus “abnormal” cells. The input of the Gradient Boosting Classifier is the 6-d vector from the six classification networks, and the output is ground-truth knowledge of whether any abnormal cells are presented in the given image patch.

\section{Results}
\label{section:result}
\subsection{Classification performance of cell patches}
To evaluate the performance of the proposed framework, we firstly test its cell type classification module (i.e. feature extraction and classification networks) on manually-identified cell patches through 5-fold cross-validation. Binary classification performance for the six cell types, as measured by Area Under the Curve (AUC), is listed in the first row in Table \ref{Table AUC}, marked by \textit{Model A}. It can be seen that classification AUC for all individual cell types is all above 0.9. Further, for a given image patch with an arbitrary number of cells belonging to same or different cell types, the proposed model can identify all the cell types and generate an exactly correct label at the accuracy of 0.722. In comparison, if the proposed model is used to classify image patches containing only a single cell (second row in Table 1, marked by \textit{Model A*}), it can achieve an overall classification accuracy of 0.932, which outperforms the accuracy reported in our previous work (0.893) \cite{Xu2017A}. The evaluation results indicate that the proposed classification method possess superior effectiveness than our previous work.

\begin{table*}[!htbp]
\centering
\small
\caption{ROC-AUC score of six classifiers with different experimental settings. \textit{Model A}: mixed dataset train, mixed dataset test. \textit{Model A*}: mixed dataset train, single-cell dataset test. \textit{Model B}: single-cell dataset train, mixed dataset test.}
\label{Table AUC}
\renewcommand{\arraystretch}{1}
\begin{tabular}{|m{1.6cm}|m{1cm}|m{1cm}|m{1cm}|m{1cm}|m{1cm}|m{1cm}|m{1.2cm}|}
\hline
\multirow{2}*{Experiment} & \multicolumn{6}{c|}{AUC} & \multirow{2}*{Accuracy}\\
\cline{2-7}
~ & Oval + Disc & Elon + Sick & Reti & Gran & Echi & Stom & ~ \\
\hline  
\textit{Model A}& 0.971 & \textbf{0.943} & \textbf{0.967} & \textbf{0.933} & \textbf{0.985} & \textbf{0.908} & \textbf{0.722} \\
\hline
\textit{Model A*} & 0.995 & 0.994 & 1.000 & 0.998 & 0.999 & 0.998 & 0.932\\
\hline 
\textit{Model B}& \textbf{0.974}  & 0.891 & 0.952 & 0.819 & 0.935 & 0.671 & 0.649 \\
\hline 
\end{tabular}
\end{table*}

In order to investigate whether the current classification module benefits from the extra multi-instance training samples, we further train a same set of six classification networks with only single-cell image patches. Its classification performance on the mixed dataset with both single and multi-cell patches is listed in the third row in Table \ref{Table AUC}, marked by \textit{Model B}. Overall classification accuracy of Model B decreases dramatically comparing with \textit{Model A} (0.649 versus 0.722). While it achieves higher accuracy for classifying “Oval+Disc” cell type (which contains the largest number of samples), for all the other five cell types its performance is lower. This comparison shows the great significance of adding multiple cell patches into training samples. 

\begin{figure}[ht]
\centering
\includegraphics[width=\textwidth]{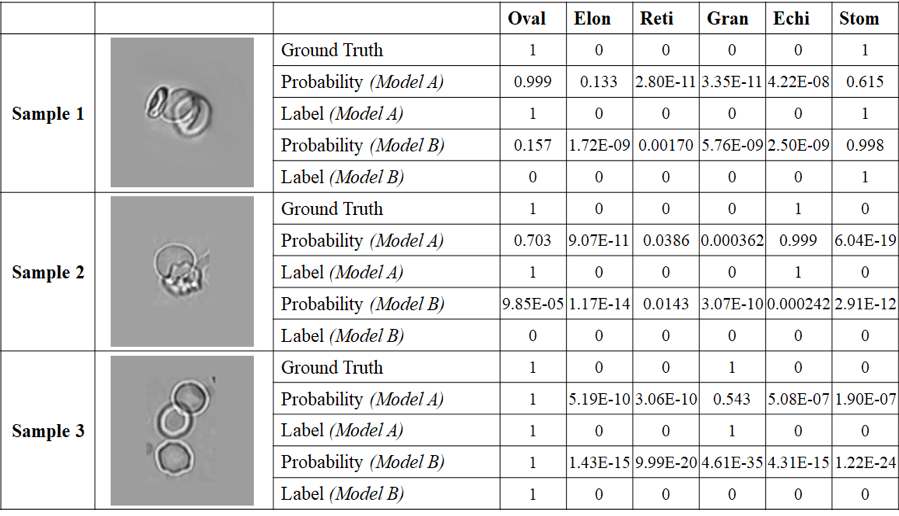}
\caption{Sample image patches containing multiple cell types and their predicting probability and label generated from \textit{Model A} and \textit{Model B}. They are all correctly classified by \textit{Model A} yet incorrectly classified by \textit{Model B}.}
\label{figure:sample of model A and B}
\end{figure}

Several sample cases where \textit{Model A} (i.e. network used in the proposed framework) makes correct classification while \textit{Model B} (single-cell network) fails are visualized in Table.\ref{figure:sample of model A and B}. For image patch “Sample 1” which contains cell types of “Oval” and “Stomatocyte”, \textit{Model A} correctly identify both cell types (with predicting probability of 0.999 and 0.615), while \textit{Model B} classify the patch as only “Stomatocyte” (with predicting probability of 0.998). For image patch “Sample 2” which contains cell types of “Oval” and “Echinocytes”, \textit{Model A} correctly identify both cell types (with predicting probability of 0.703 and 1), while \textit{Model B} predicts no label for the patch (i.e. outputs from all six networks are lowered than the threshold). For image patch “Sample 3” which contains cell types of “Oval” and “Granular”, \textit{Model A} correctly identify both cell types (with predicting probability of 1 and 0.542), while \textit{Model B} classify the patch as only “Oval” (with predicting probability of 1). It can be found that for image patches containing multiple cell types, \textit{Model B} will either predict only one label or no label at all, while \textit{Model A} can identify all the correct cell types. The result shows that only by adding multi-cell data into the training samples, the classification network can learn how to handle them accurately.

Finally, we train the Gradient Boosting Classifier from outputs of the six classification networks for patch-wise SCD testing. The proposed Gradient Boosting Classifier achieves an average accuracy of 85.1\% through 5-fold cross-validation, indicating that for a given image patch with arbitrary number of cells belonging to same or different cell types, the classifier can determine whether there is at least one abnormal cell at high accuracy.

In order to investigate whether the current classification module benefits from the extra multi-label training samples, we further train a same set of six classification networks with only single-cell image patches. Its classification performance on the mixed dataset with both single and multi-cell patches is listed in the third row in Table 1, marked by \textit{Model B}. Overall classification accuracy of \textit{Model B} decreases dramatically comparing with \textit{Model A} (0.649 versus 0.722). While it achieves higher accuracy for classifying "Oval+Disc" cell type (which contains the largest number of samples), for all the other five cell types its performance is lower. 

\subsection{Automatic analysis of full-scale microscopic images}

By applying the Faster-RCNN module of the proposed framework on the full-scale input image, we can automatically obtain bounding-box of potential cells and the corresponding image patches for later classification analysis.

\begin{figure}[h]
\centering
\includegraphics[width=0.8\textwidth]{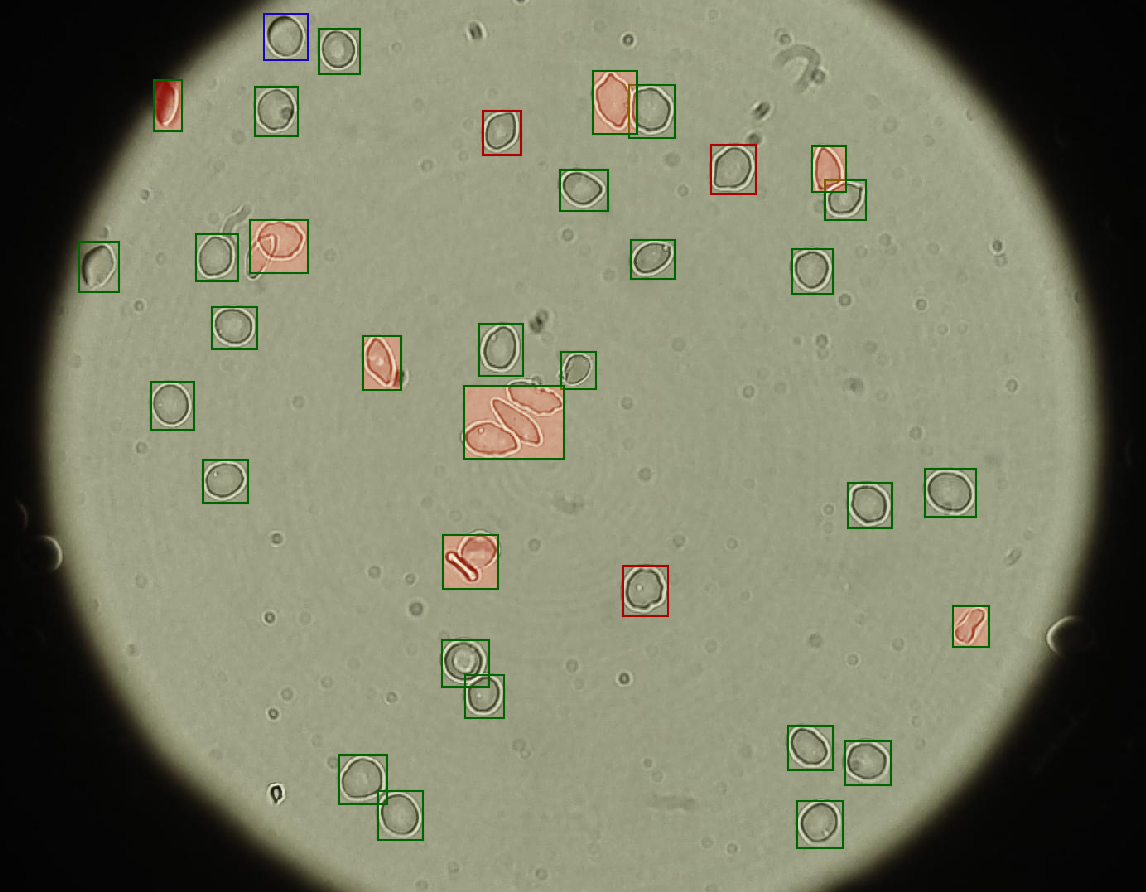}
\caption{Cell detection, classification and diagnosis result of the framework. Cell regions (i.e. image patches) are highlighted by colored boxes. Patches missed by the detection network (Faster-RCNN) are colored in blue. Patches both successfully detected and classified are colored in green. Patches that are successfully detected yet mis-classified are colored in red. Red masked regions highlight the presence of abnormal cell types, correctly detected and classified by the proposed framework.}
\label{figure:whole_result}
\end{figure}

For Faster-RCNN, our evaluation metric, average precision (AP), is the area under precision-recall curve. For every detected cell region, there is a model score generated between 0 and l, showing the level of confidence in this region. If we set a threshold, making regions with scores bigger than this threshold positive samples and vice versa, then there will be a precision-recall coordinate for all test cell regions. By varying this threshold from 0 to 1, we get a precision-recall curve and the area under this curve is called average precision (AP). In our experiment, AP on the test data is 0.899. A sample cell detection and classification result are shown in Figure~\ref{figure:whole_result}. The sample result illustrates that our proposed framework is capable of performing fully automatic cell detection and classification from raw image input, achieving end-to-end image-based SCD testing, and readily usable in real practice.
 
\section{Conclusions and Discussion}
\label{section:conclusion}
\subsection{Discussion}
There have been several successful methods for cell segmentation; however, our framework doesn’t adopt these segmentation-based methods because they don’t satisfy our classification requirements. In order to compare the performance between these methods and our model, we design a comparison experiment. 

\begin{figure}[ht]
\centering
\includegraphics[width=0.8\textwidth]{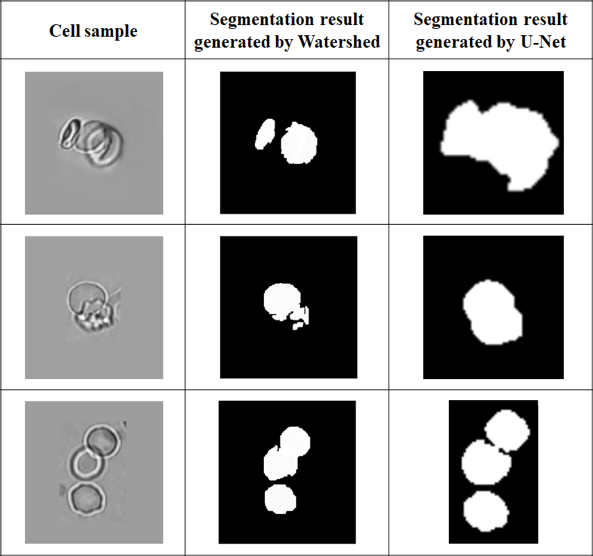}
\caption{Segmentation result generated by Watershed and U-Net.}
\label{figure:U-Net}
\end{figure}

We use watershed algorithm to perform cell segmentation on the image samples showed in section 3.1. The local result is shown in Figure~\ref{figure:U-Net}. Clearly, this unsupervised algorithm tends to merge overlapping cells and touching cells into one connected component, which contradicts with our segmentation aim. And even though it seems that Watershed can separate overlapping cells in the first sample, the segmentation result shows that one of overlapping cells is missed. 
 
On the other hand, we also use deep learning-based method for segmentation. The U-Net architecture has been shown to offer a precise localization for image semantic segmentation. It has been implemented on microscopic red blood cell images for red blood cell detection \cite{zhang2018rbc}. We use it to perform cell segmentation on the whole microscopic images which contain samples shown in Figure~\ref{figure:sample of model A and B} and cut the segmentation of the cells in cell samples from the output images which is shown in Figure~\ref{figure:U-Net}. The result shows that U-Net architecture also can’t separate overlapping and touching cells.

The segmentation result shows that it is extremely hard to separate overlapping cells and touching cells. To solve this problem, we propose a multi-cell detection and multi-label classification method which can classify multi-cell patches directly and avoid the difficult task of separating overlapping cells and touching cells.

\subsection{Conclusion}
In this work, we propose a deep learning-based framework to perform automatic cell detection and classification from RBC microscopic images. The framework is specifically designed to solve complex imaging scenario involving multi-label classification problem, where cells in the input image can be touching or overlapping with each other and cannot be separated. Experimental results show that the classification networks utilizing transfer learning scheme can achieve better performance than baseline models and previous works, deal with more complex cell imaging conditions and partially address the highly challenging multi-label classification problem. Testing results on full-scale raw microscopic image input show high robustness of the proposed framework and its potential usefulness in clinical practice.






\small
\bibliographystyle{unsrt}
\bibliography{refs}

\end{document}